\documentclass[%
reprint,
 amsmath,amssymb,
 aps,
]{revtex4-1}

\usepackage{tikz}
\usepackage{graphicx}% Include figure files
\usepackage{dcolumn}% Align table columns on decimal point
\usepackage{bm}% bold math
\usepackage{libertine}
\usepackage{color}
\usepackage{hyperref}% add hypertext capabilities

\newcommand*\stoneblack[1]{\tikz[baseline=(char.base)]{%
            \node[shape=circle,fill=black!100,font=\fontsize{9}{10}\selectfont,draw,inner sep=0.5pt] (char) {\color{white}#1};}}           
\newcommand*\stonewhite[1]{\tikz[baseline=(char.base)]{%
            \node[shape=circle,fill=white!100,font=\fontsize{9}{10}\selectfont,draw,inner sep=0.5pt] (char) {\color{black}#1};}}            
\newcommand*\stoneblacksmall[1]{\tikz[baseline=(char.base)]{%
            \node[shape=circle,fill=black!100,font=\fontsize{7.5}{10}\selectfont,draw,inner sep=0.3pt] (char) {\color{white}#1};}}           
\newcommand*\stonewhitesmall[1]{\tikz[baseline=(char.base)]{%
            \node[shape=circle,fill=white!100,font=\fontsize{7.5}{10}\selectfont,draw,inner sep=0.3pt] (char) {\color{black}#1};}}
     
\newsavebox{\blackstoneone}
\newsavebox{\whitestonetwo}
\newsavebox{\blackstonethree}
\newsavebox{\whitestonefour}
\newsavebox{\blackstonefive}
\sbox\blackstoneone{\stoneblacksmall{1}}
\sbox\whitestonetwo{\stonewhitesmall{2}}
\sbox\blackstonethree{\stoneblacksmall{3}}
\sbox\whitestonefour{\stonewhitesmall{4}}
\sbox\blackstonefive{\stoneblacksmall{5}}

\begin{document}

\title{Quantum \emph{Go} Machine}

\author{ Lu-Feng Qiao$^{1,2}$, Jun Gao$^{1,2}$, Zhi-Qiang Jiao$^{1,2}$, Zhe-Yong Zhang$^{1,2}$, Zhu Cao$^{3}$, Ruo-Jing Ren$^{1,2}$, Chao-Ni Zhang$^{1,2}$, Cheng-Qiu Hu$^{1,2}$, Xiao-Yun Xu$^{1,2}$, Hao Tang$^{1,2}$, Zhi-Hao Ma$^{4}$}
\author{Xian-Min Jin$^{1,2}$}
\email{xianmin.jin@sjtu.edu.cn}

\affiliation{$^1$Center for Integrated Quantum Information Technologies (IQIT), School of Physics and Astronomy and State Key Laboratory of Advanced Optical Communication Systems and Networks, Shanghai Jiao Tong University, Shanghai 200240, China}
\affiliation{$^2$CAS Center for Excellence and Synergetic Innovation Center in Quantum Information and Quantum Physics, University of Science and Technology of China, Hefei, Anhui 230026, China}
\affiliation{$^3$Key Laboratory of Advanced Control and Optimization for Chemical Processes of Ministry of Education, East China University of Science and Technology, Shanghai 200237, China}
\affiliation{$^4$School of Mathematical Sciences, Shanghai Jiao Tong University, Shanghai 200240, China}

\maketitle

\textbf{\emph{Go} has long been considered as a testbed for artificial intelligence. By introducing certain quantum features, such as superposition and collapse of wavefunction, we experimentally demonstrate a quantum version of \emph{Go} by using correlated photon pairs entangled in polarization degree of freedom. The total dimension of Hilbert space of the generated states grows exponentially as two players take turns to place the stones in time series. As nondeterministic and imperfect information games are more difficult to solve using nowadays technology, we excitedly find that the inherent randomness in quantum physics can bring the game nondeterministic trait, which does not exist in the classical counterpart. Some quantum resources, like coherence or entanglement, can also be encoded to represent the state of quantum stones. Adjusting the quantum resource may vary the average imperfect information (as comparison classical \emph{Go} is a perfect information game) of a single game. We further verify its non-deterministic feature by showing the unpredictability of the time series data obtained from different classes of quantum state. Finally, by comparing quantum \emph{Go} with a few typical games that are widely studied in artificial intelligence, we find that quantum \emph{Go} can cover a wide range of game difficulties rather than a single point. Our results establish a paradigm of inventing new games with quantum-enabled difficulties by harnessing inherent quantum features and resources, and provide a versatile platform for the test of new algorithms to both classical and quantum machine learning.} 
\bigskip

\section*{Introduction}
\emph{Go} has represented a typically challenging game for artificial intelligence due to its tremendous search space. For other classical board games such as chess and checkers, researchers have trained the programs to exceed professional players by using brute force tree-search combined with human expertise in early days\cite{Schaeffer2000}. However, brute force tree-search cannot deal with the game of \emph{Go}; enumeration of all possible moves seems like an impossible mission for computers. To solve this problem, the researchers adopted the Monte Carlo tree search method in programming, which made programs achieving a strong amateur player's level\cite{Kocsis2006, Coulom2006, Browne2012}. In 2016, AlphaGo shocked the world by beating Lee Sedol in a five-game match using deep neural networks \cite{Silver2016}. Later, Google announced that a new program AlphaGo Zero which based solely on reinforcement learning defeated AlphaGo with 100:0 after a short time self-playing \cite{Silver2017}.

As one of the most complex board games in terms of possible game states, \emph{Go} is actually not the hardest game to tackle for machine learning algorithms. The difficulty not only depends on the complexity of the game (state-space complexity or game-tree complexity\cite{Allis1994}) but also is highly related to the features and strategies of the game. The games are therefore classified in light of their features \cite{Pumperla2019}: deterministic/nondeterministic and perfect/imperfect information. All the gambling games are nondeterministic games, as shuffling cards or rolling dice brings the randomness into these games. In imperfect information games, all players can only access a part of game states. 

\emph{Go} is a deterministic (the course of a game is only determined by players' decisions) and perfect information game (both players can see all the stones on the board, no player has private information about the game state that the other player does not know). After Alpha \emph{Go}, the community moved interest to nondeterministic and imperfect information games, like Poker, Mahjong and even video games like StarCraft, Dota2 \cite{Moravcik2017,Brown2018,Brown2019,Vinyals2017,Semenov2016}. In these games, the players need to guess what other players know, and consider the factors induced by the uncontrolled randomness, which makes the games much more difficult to solve\cite{Frank1998}. To master these games provides a benchmark at a higher level for machine learning algorithms. 

We find that the nondeterministic and imperfect information traits of games perfectly suit the inherent features of quantum physics, for example, quantum-enabled intrinsic randomness. To realize a quantum version of game \emph{Go}, we employ entangled photons associated with the built-in superposition and randomness to simulate the quantum stones, which can occupy two places simultaneously instead of only one intersection. Apart from the nondeterministic feature, the imperfect information feature can also be introduced to quantum \emph{Go} by using non-maximally entangled states, where the biased correlation can keep the private information to the other player until the state is measured. The nondeterministic and imperfect degree of the game can be tuned by engineering the quantum entanglement. Thus, quantum \emph{Go} is different from all other nondeterministic and imperfect information games since it can cover a wide range of game difficulties rather than a single point, which gives a wide range of benchmarks for artificial intelligence. 

The phenomenon that superposition states can exist in quantum systems\cite{Monroe1996}  is a significant feature in quantum physics and enables many applications that are impossible in the classical world. A crossover between various modern sciences and quantum mechanical laws has stimulated many promising technologies such as quantum communication\cite{Bennett1984, Jin2010, Pan2012, Lo2014}, quantum computation\cite{Shor1994, Grover1997}, and quantum machine learning\cite{Biamonte2017}. In classical machine learning, researchers tend to use the classical board and card games as the testbeds because these games provide closed worlds with specific and simple rules\cite{Gelly2014} as well as a clear benchmark. While quantum versions of various machine learning algorithms have been proposed\cite{Lloyd2014, Rebentrost2014, Lloyd2013} and experimentally implemented\cite{Melnikov2018, Cai2015, Li2015, Gao2018}, a testbed for these algorithms remains elusive. In this work, we propose a quantum version of \emph{Go} that could be an excellent candidate as a testbed for both classical and quantum machine learning. 

The paper is organized as following, we first introduce the basic rule of quantum \emph{Go}, then experimentally demonstrate the game. We get the high-quality quantum stones using polarization-entangled photons and we employ a time-of-flight storage module to collect massive time series data. We demonstrate different nondeterministic and imperfect degrees of the game by tuning the quantum entanglement. Finally, we present a demo for the data we collected by showing a real Kifu in finite moves.

\section*{The features of quantum \emph{Go}}
In the quantum version of game \emph{Go}, the first modification of classical \emph{Go} (see Methods for the rules of classical Go) is that the players can put a quantum stone at two intersections simultaneously each turn based on the superposition principle of quantum mechanics. 
The quantum stones will only occupy the intersections, but they will not reduce the liberties (when a stone has no liberty it will be captured) of neighbors' stones. It means that one can not capture the other's stones with quantum stones.

Another unique feature of quantum \emph{Go} is the quantum-like collapse measurement rule, when one player decides to place a quantum stone adjacent to the existed stones, the collapse process takes place. 
In the quantum world, the measurement induces a superposition state collapse to a certain classical state. Such a measurement may not be done on purpose, as long as the state interacts with the environment that makes it possible for one to extract the information of the states in principle, no matter whether or not there exists a conscious being actually reaching that information. 

The same idea is transferred to quantum \emph{Go}, where we define the directly adjacent (up, down, left, and right) intersections of a quantum stone as its detectable area. The quantum stones will not interact with empty intersections until another stone goes into its detectable area, making the positional information of the quantum stone determined.
After the collapse process, the quantum stone will be determinately settled in one of the two intersections, and become a classical stone. 

\begin{figure}[!b]
\centering
\includegraphics[width=0.98\columnwidth]{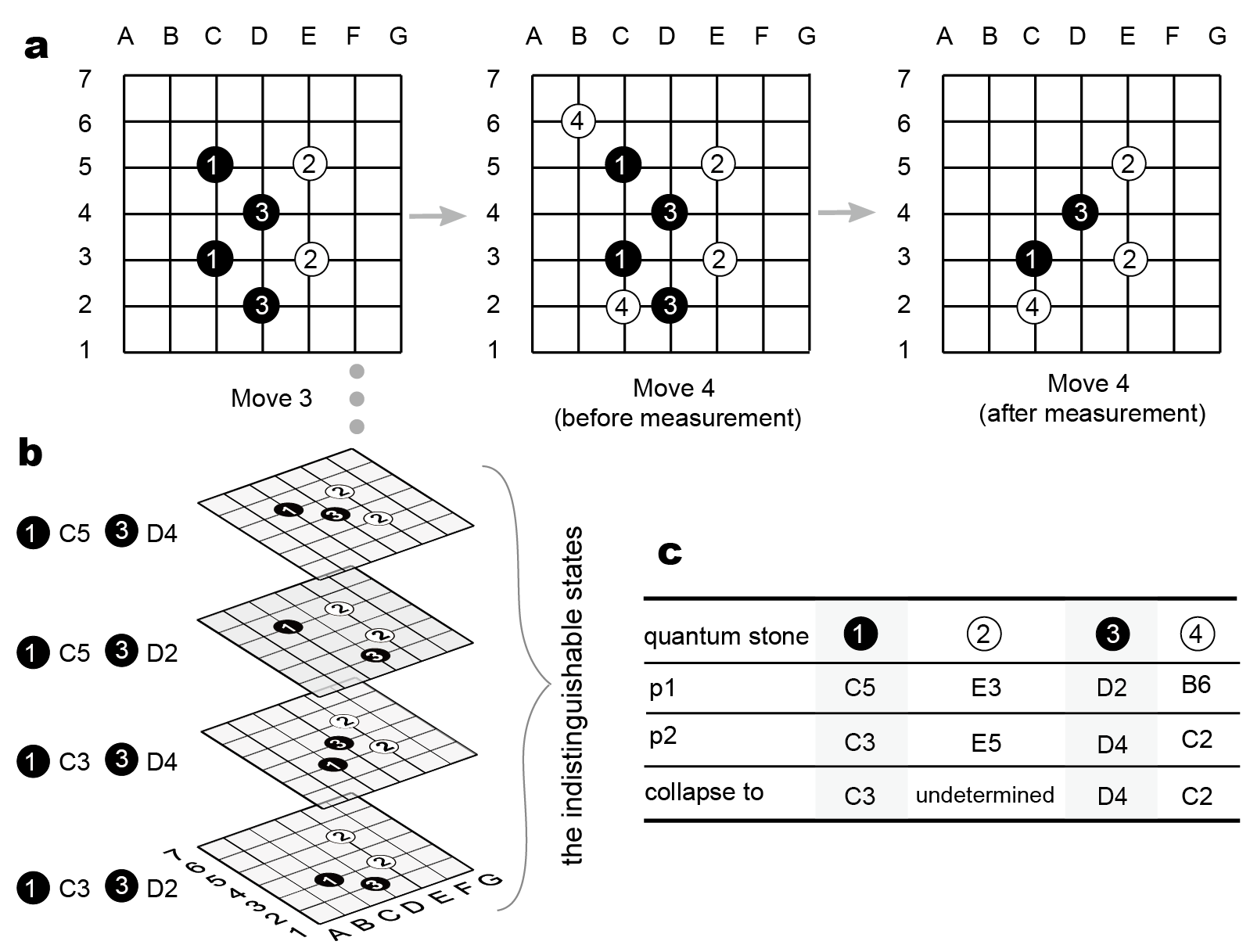}\\
\caption{\textbf{Placing stones and collapse measurement.} \textbf{a}, An example of the first four moves of a game. In Move 3, there are 3 quantum stones on the board. In Move 4, as \usebox{\whitestonefour} is placed next to \usebox{\blackstoneone} and \usebox{\blackstonethree}, the detectable area of these three quantum stones contains other stones, which causes the collapse measurements. After the measurements, \usebox{\blackstoneone}, \usebox{\blackstonethree} and \usebox{\whitestonefour} become classical stones. \textbf{b}, The indistinguishable states for the white player making the decision to place \usebox{\whitestonefour}. In the manner of imperfect information games, these states will be put into the same information set of the white player. \textbf{c}, The choices of $p_1$ and $p_2$ corresponding to the two intersections of quantum stone and the collapse measurement results are recorded in the table.}
\label{Figure1}
\end{figure}

\begin{figure*}
\centering
\includegraphics[width=1.98\columnwidth]{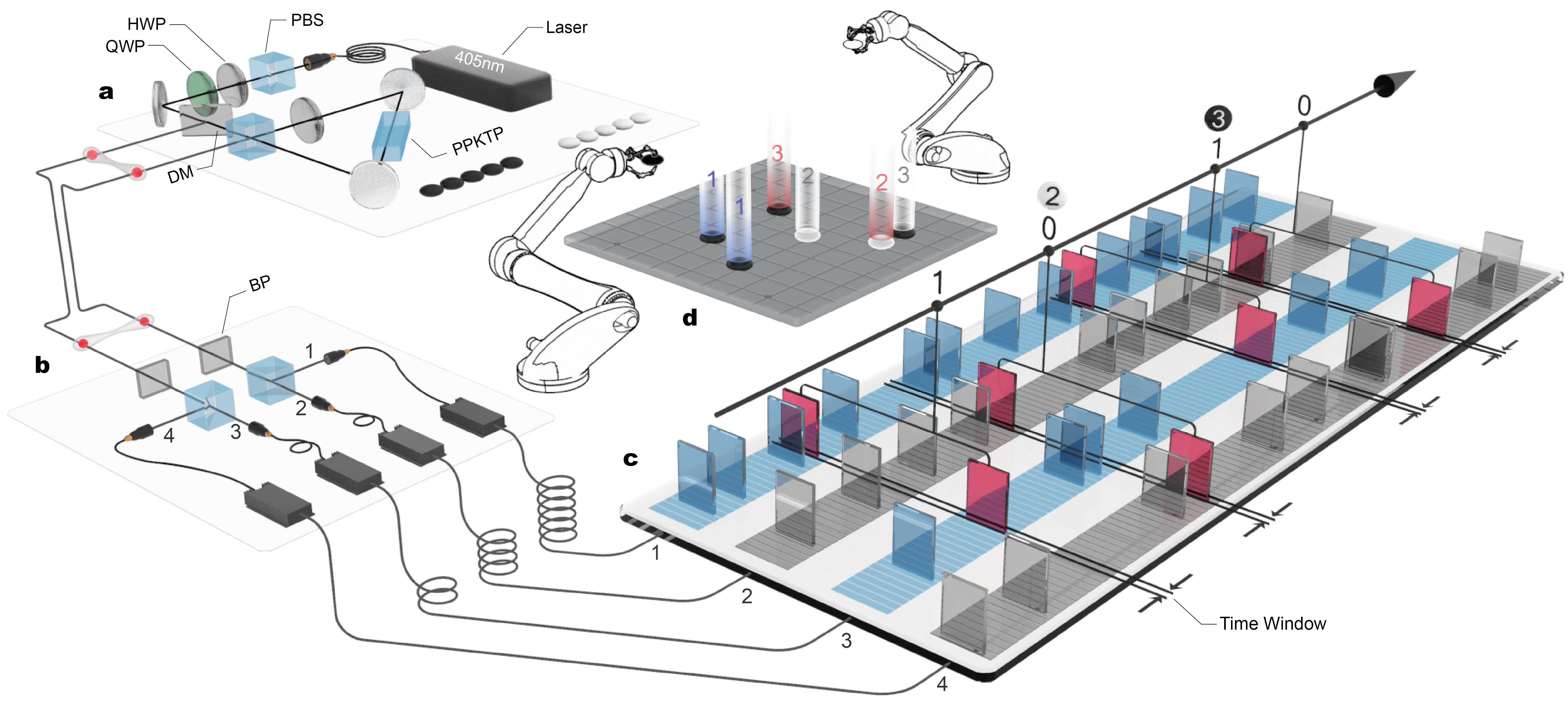}\\
\caption{\textbf{Sketch of quantum \emph{Go} machine.} 
\textbf{a}, Experimental setup of the quantum stone box. The generated photon pairs can be tuned to maximally entangled states, non-maximally entangled states and product states to behave as different quantum stones, see Methods. \textbf{b}, The collapse measurement module. After the photons come into this module, they will be measured by the polarizing beam splitter (PBS) then the quantum state collapses to path 1 and 3 (or path 2 and 4). Four single photon detectors transfer the photon signals to electronic signals. \textbf{c}, The time-of-flight storage module. Four output channels from the collapse measurement module will be guided into this module. The collapse result information of each pair of the entangled photons can be acquired after setting a proper coincidence time window, and recorded as an effective stored state in the time series data. We encode the signals coincidence in Channel 1 and 3 as ``1", and Channel 2 and 4 as ``0''. \textbf{d}, Sketch of playing quantum \emph{Go} with the quantum stones from the time series data. Two robot arms represent the two agents who help to execute the game of quantum \emph{Go} together. They pick the quantum stones from the quantum stone box alternately and put every stone onto two intersections of the virtual board. When a quantum stone is put on an intersection that has neighbors, the game will get the collapse results from the time series data with a backdated measurement in the collapse measurement module.
}
\label{Figure2}
\end{figure*}

A quantum stone will occupy two intersections. The player who places the quantum stone would choose one intersection as position $p_1$ and the other as position $p_2$. The information of choices is sent to the referee or a judgment system while the information is kept secret to the other player until the stone is measured. Each quantum stone can be expressed as:
\begin{equation}
\arrowvert{\psi}\rangle=a_1\arrowvert{1}\rangle_{p_1}\arrowvert{0}\rangle_{p_2}+a_2\arrowvert{0}\rangle_{p_1}\arrowvert{1}\rangle_{p_2},
\label{Equantion1}
\end{equation} 
where $\arrowvert{1}\rangle_{p_1}\arrowvert{0}\rangle_{p_2}$ represents a stone on the intersection $p_1$ not on $p_2$, and vice versa. $\vert{a_1}\vert^{2}$ and $\vert{a_2}\vert^{2}$ are proportional to the probability to collapse onto $p_1$ and $p_2$ respectively. 

Fig.\ref{Figure1}\textbf{a} shows the first four moves of a game. In move 4, \stonewhite{4} is placed to be adjacent to \stoneblack{1} and \stoneblack{3}, which causes these three stones to be measured. In the game state of move 3, there are two black quantum stones, which can collapse to 4 different game states (Fig.\ref{Figure1}\textbf{b}). These 4 states are indistinguishable for the white player to make the strategy of placing \stonewhite{4}. The player's choices of ${p_1}$ and ${p_2}$ are recorded (Fig.\ref{Figure1}\textbf{c}), and making the game an imperfect information game (see detail in Methods).

In addition, if ${\vert{a_1}\vert}^{2}={\vert{a_2}\vert}^{2}$, the choice information of $p_1$ and $p_2$ will not give one player more information, in other words, the game has no private information. Meanwhile, if ${\vert{a_1}\vert}^{2}\neq{\vert{a_2}\vert}^{2}$, the game will have the private information that makes the game an imperfect information game.

\bigskip
\bigskip

\section*{Experimental implementation and Results}

\begin{figure*}
 \centering
\includegraphics[width=1.98\columnwidth]{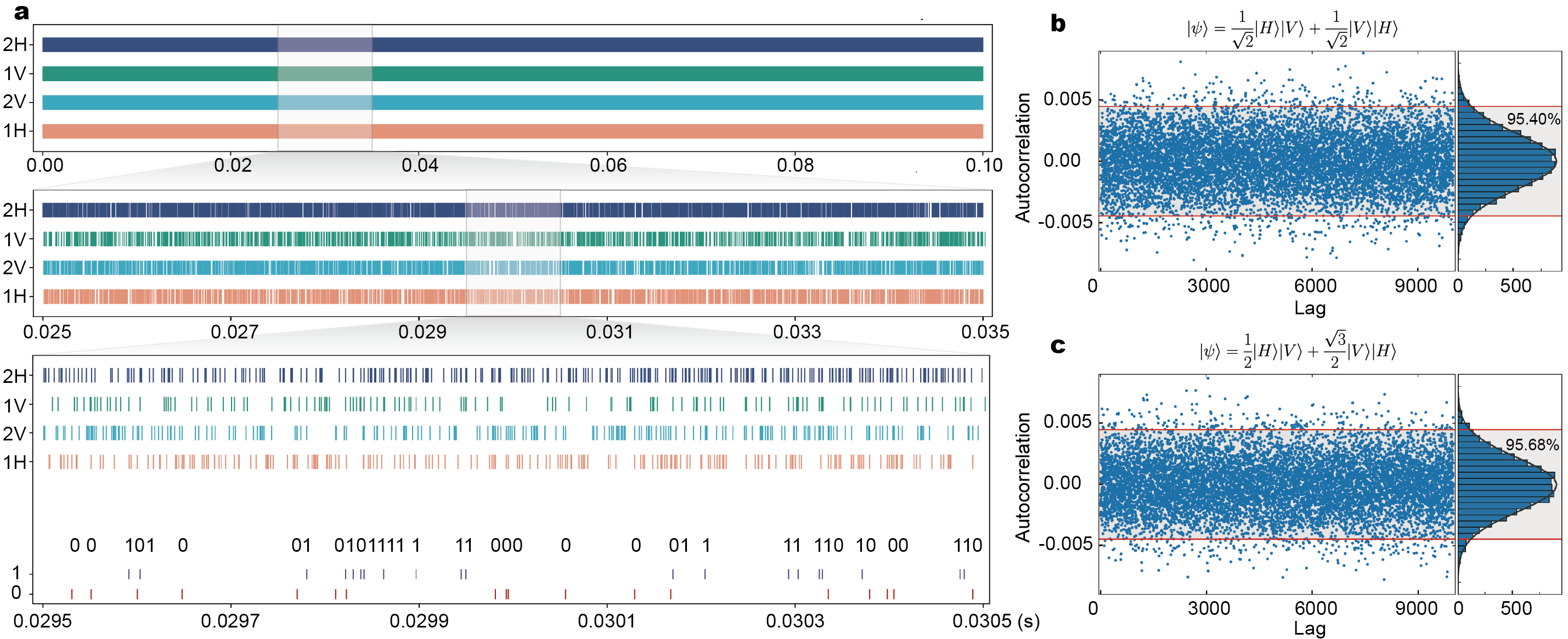}\\
\caption{\textbf{Measured time series data of maximally entangled states and autocorrelation measurements.} \textbf{a}, A section of the time series data from the four channels within 0.1 seconds. The insets below show the quantum stones by further zooming the time scale of 0.01 seconds and 0.001 seconds. \textbf{b}-\textbf{c}, The experimental results of autocorrelation measurement. The lag is applied from 1 to 10000 for the time series data of 200000 quantum stones, for maximally entangled states (\textbf{b}) and non-maximally entangled states (\textbf{c}). The red line gives the $95\%$ confidence bounds. The insets on the right give the statistics of the number of different autocorrelation values. The quantity within the $95\%$ confidence bound is $95.40\%$ and $95.68\%$ respectively. }
 \label{Figure3}
\end{figure*}

\subsection{Entangled photons act as quantum stones}

In our experiment, we use 2-qubit correlated photon pairs to be the quantum stones. As the two players take turns to place the quantum stones, the total dimension of Hilbert space of the generated states grows exponentially. Since the generated state is a tensor product state, we can use time series photon pairs to simulate the process
\begin{equation}
\begin{split}
\left|\Psi_{1}\right\rangle =\bigotimes_{n=1}^{N}\left|\psi\right\rangle _{\tau_{2n-1}}, \left|\Psi_{2}\right\rangle =\bigotimes_{n=1}^{N}\left|\psi\right\rangle _{\tau_{2n}},
\end{split}
\label{Equantion2}
\end{equation} 
where $\tau$ together with index $n(n\in\mathbb{N}^+)$ indicate the discrete time of placing the stones, $\left|\Psi_{1}\right\rangle$ and $\left|\Psi_{2}\right\rangle$ represent the states for two players respectively, and the total state can be expressed as $\left|\Psi\right\rangle =\left|\Psi_{1}\right\rangle \otimes\left|\Psi_{2}\right\rangle$.
\textcolor{blue}{}

In our experiment, the state
$\arrowvert{\psi}\rangle_{\tau_{n}} (n\in\mathbb{N}^+)$ for the two players are realized by a pair of entangled photons that are spectrally indistinguishable. Unlike the classical counterpart, all quantum \emph{Go} stones can be generated in one stone box and are identical before measurement. As shown in Fig.\ref{Figure2}\textbf{a}, we introduce details of the preparation of the quantum stone box, where polarization encoded photons are made to act as the quantum stones (see Methods). The state of each entangled photon pair can be expressed as
 \begin{equation}
\arrowvert{\psi}\rangle_{\tau_{n}}=\cos\theta\arrowvert{H}\rangle_{1}\arrowvert{V}\rangle_{2}+e^{i\phi}\sin\theta\arrowvert{V}\rangle_{1}\arrowvert{H}\rangle_{2}.
\label{Equantion3}
\end{equation}  

For convenience, we often omit the subscript of the states.
Since Eq.(\ref{Equantion3}) and Eq.(\ref{Equantion1}) are very similar in form, it would be natural to map the superposition of the quantum stones' locations on the board onto the superposition of $\arrowvert{H}\rangle\arrowvert{V}\rangle$ and $\arrowvert{V}\rangle\arrowvert{H}\rangle$ for entangled photons in the Hilbert spaces. 

In each turn, the players can decide their possible moves. The positions of the stones are recorded in a virtual game board (e.g., a computer terminal). When a player places a stone that causes the collapse measurement subsection, all the involved entangled photons are measured by two polarizing beam splitters (Fig.\ref{Figure2}\textbf{b}). There is a probability of ${\vert{\cos\theta}\vert}^{2}$(${\vert{\sin\theta}\vert}^{2}$) that photon 1 reflects to path1 (transmits to path2) and photon 2 transmits to path3 (reflects to path4), which will determine the quantum stone collapse onto intersection $p_1$($p_2$). That is, we can get a definite statistical possibility of the collapse measurement result for a determined state. However, for each certain photon pair, whether it comes out to path $1\&3$ or path $2\&4$ after the measurement is unpredictable, no matter how advanced the experimental apparatus is. This unpredictable nature brings the intrinsic nondeterministic feature to quantum \emph{Go}. 

Once the collapse process finishes, the game moves back to the normal turn. In this way, we are able to simulate a large-dimensional tensor product state by picking entangled states and collapse measurements in a time series fashion. This scheme requires that the entangled states should be identical and stable. Since the wavelength of the down-converted photons is tunable via the temperature of the pumped crystal, we manage to lock the temperature within $\pm5mK$ by using a PID controller. Two additional bandpass filters are used to guarantee the entangled photons spectrally indistinguishable. The brightness of the entanglement source is high enough so that the statistical fluctuation of the photon counts is negligible.

\begin{figure*}
 \centering
 \includegraphics[width=1.98\columnwidth]{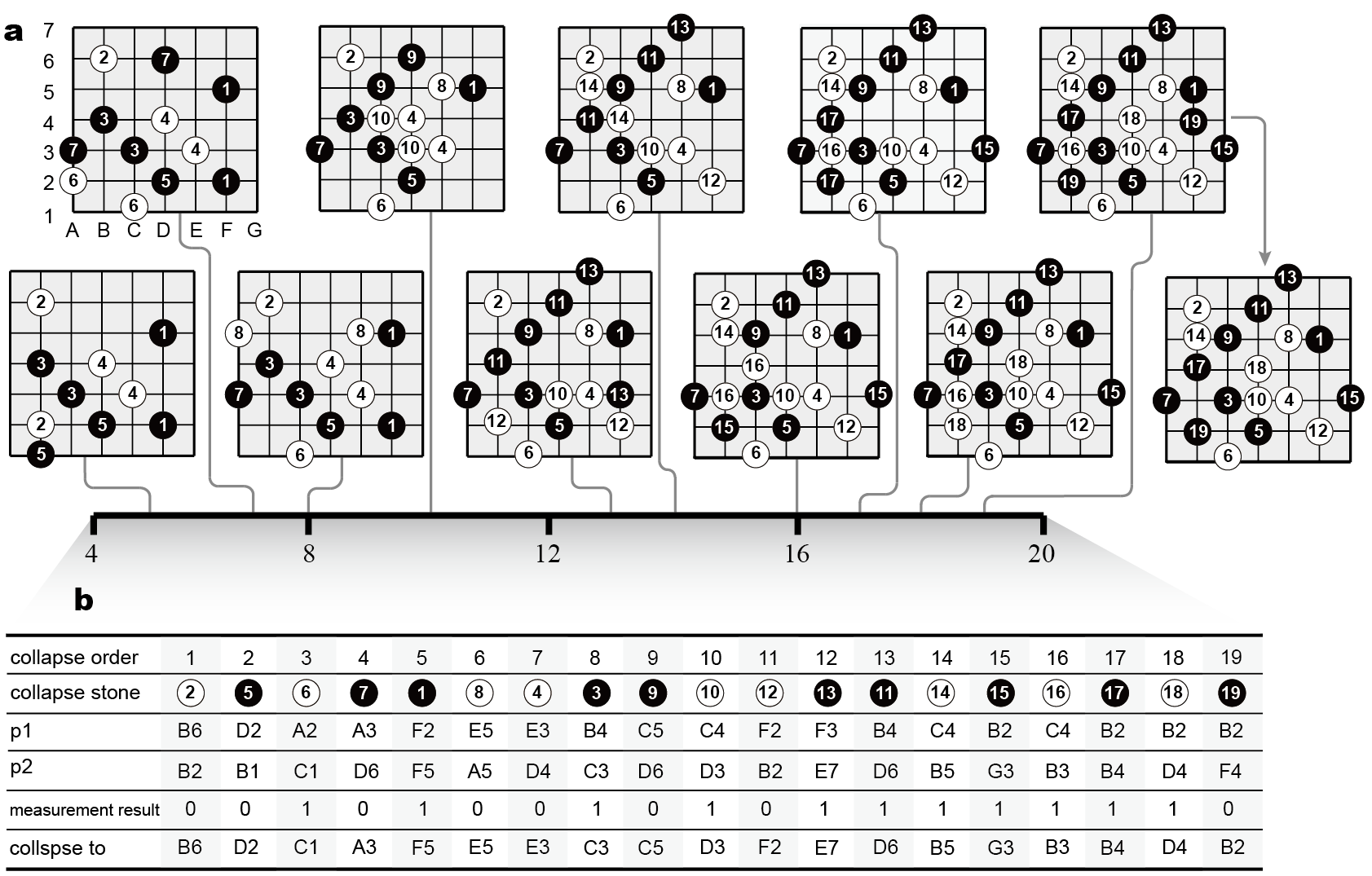}\\
 \caption{\textbf{Experimental demonstration of quantum \emph{Go} on a $7\times7$ board.} \textbf{a}, Selected Kifu played on a $7\times7$ board. On the leftmost board, there are five quantum stones. \usebox{\blackstonefive}, which is set adjacent to \usebox{\whitestonetwo}, will cause the quantum stone 2 and 5 to collapse. The first several moves are shown on the board. Specifically, when placing the 5th, 7th, 8th, 10th, 13th, 14th, 16th, 17th, 18th and 19th quantum stones, the collapse measurements are triggered by the moves. \textbf{b}, The table shows the specific moves, collapse measurement results and the induced intersections according to the time series data and the players' choices.}
 \label{Figure4}
\end{figure*}

\subsection{Experiment results}

We obtain the collapse information of each entangled photon pair {(see Methods) other than the statistical result in a period of time in our experiment.  As shown in Fig.\ref{Figure2}\textbf{c}, the outputs of the measurement module are guided into 4 channels of the time-of-flight storage module. While the output of each single photon detector includes noises due to the dark counts of the detector itself and the background from the environment, the signal of entangled photons can still be well identified by coincidence measurement with negligible error. The collapse information of each pair of entangled photons can be acquired after setting a proper coincidence time window (in our experiment, $2 ns$). The coincidence events of Channel 1 and Channel 3 (Channel 2 and Channel 4) would be coded as ``1''(``0").  

The entangled photons are continuously produced, which can compose an infinite 0/1 sequence as time flows. The generated sequence stored as time series data is the random resource for the game of quantum \emph{Go}. A sketch of playing quantum \emph{Go} with the quantum stones is shown in Fig.\ref{Figure2}\textbf{d}.
In the experiment, terabyte of time series data can be acquired in dozens of  hours. The data sampled in 1 hour includes about $N =10^8$ pairs of entangled states with a total dimension of Hilbert space up to $2^{10^8}$, which could support a game with $\lfloor\sqrt{N}\rfloor\times\lfloor\sqrt{N}\rfloor$ board and the moves up to $N$.
Fig.\ref{Figure3}\textbf{a} shows the time-labeled data of four channels in 0.1 seconds processed by time-of-flight storage module.

In order to check whether the time series data have some hidden correlations upon time, we perform the measurements of autocorrelation function. The test results are shown in Fig.\ref{Figure3}\textbf{b}-\textbf{c} and Extended Data Fig.2, which give the sample autocorrelation coefficients of the time series generated from 6 different quantum states. For a large sample of size $N$, if the time series data have no time correlation, the lagged-correlation coefficient should be normally distributed with a mean value of 0 and a variance of $1/N$\cite{Chatfield2004}.  The 95\% confidence limits are approximately represented as $r_{0.95} = 0\pm\frac{2}{\sqrt{N}}$. If a series is truly random, there is also $5\%$ chance for each lagged autocorrelation coefficient outside the  $95\%$ confidence limits. Fig.\ref{Figure3}\textbf{b-c} are the correlograms of the tested series with $N = 200000$, $r_{0.95} = \pm0.0045$. The proportions of the autocorrelation coefficients between the two red lines are 95.40\% (Fig.\ref{Figure3}\textbf{b}) and 95.68\% (Fig.\ref{Figure3}\textbf{c}) which are all larger than 95\%. It shows that the time series are not autocorrelated, that is, the new observations can not be predicted by all the past observations. 

It is quite counter-intuitive that the time series data in a biased sequence is also unpredictable. For example, if the entangled state is $\arrowvert{\psi}\rangle=\frac{1}{2}\arrowvert{H}\rangle\arrowvert{V}\rangle+\frac{\sqrt{3}}{2}\arrowvert{V}\rangle\arrowvert{H}\rangle$, ``1" is three times more than ``0" in the generated sequence. However, as shown in Fig.\ref{Figure3}\textbf{c}, the time series also passes the autocorrelation test. The collapse measurement result of each entangled pair cannot be predicted by using the past information. With the unpredictability endorsed by the inherent randomness of quantum mechanics, the quantum version of \emph{Go} behaves as an ideal nondeterministic game. For ultimate scenarios that the state $\arrowvert{\psi}\rangle_{\tau_{n}}$ are tuned to be very biased down to $\arrowvert{H}\rangle\arrowvert{V}\rangle$ or $\arrowvert{V}\rangle\arrowvert{H}\rangle$, the quantum version of \emph{Go} becomes a deterministic game, where the collapse results have been predicted by the player who places the stones but will only be exposed to the other player after the measurement.

It is inevitable to get the undesired components $\arrowvert{H}\rangle\arrowvert{H}\rangle$ and $\arrowvert{V}\rangle\arrowvert{V}\rangle$ due to the noises and multi-photon events from the entanglement source. In our experiment, the visibility is $(N_{HV}+N_{VH}-N_{HH}-N_{VV})/(N_{HV}+N_{VH}+N_{HH}+N_{VV})=99.2\%$ for the maximally entangled state \cite{Fedrizzi2007}.  The quantum state tomography\cite{James2001} is used to evaluate the
entanglement, and the concurrence\cite{Wootters1998} is up to 0.93. Once the collapse measurement gives the undesired components, the players will discard the state and retrieve a new one from the stored time series data.

\begin{figure*}
 \centering
 \includegraphics[width=1.98\columnwidth]{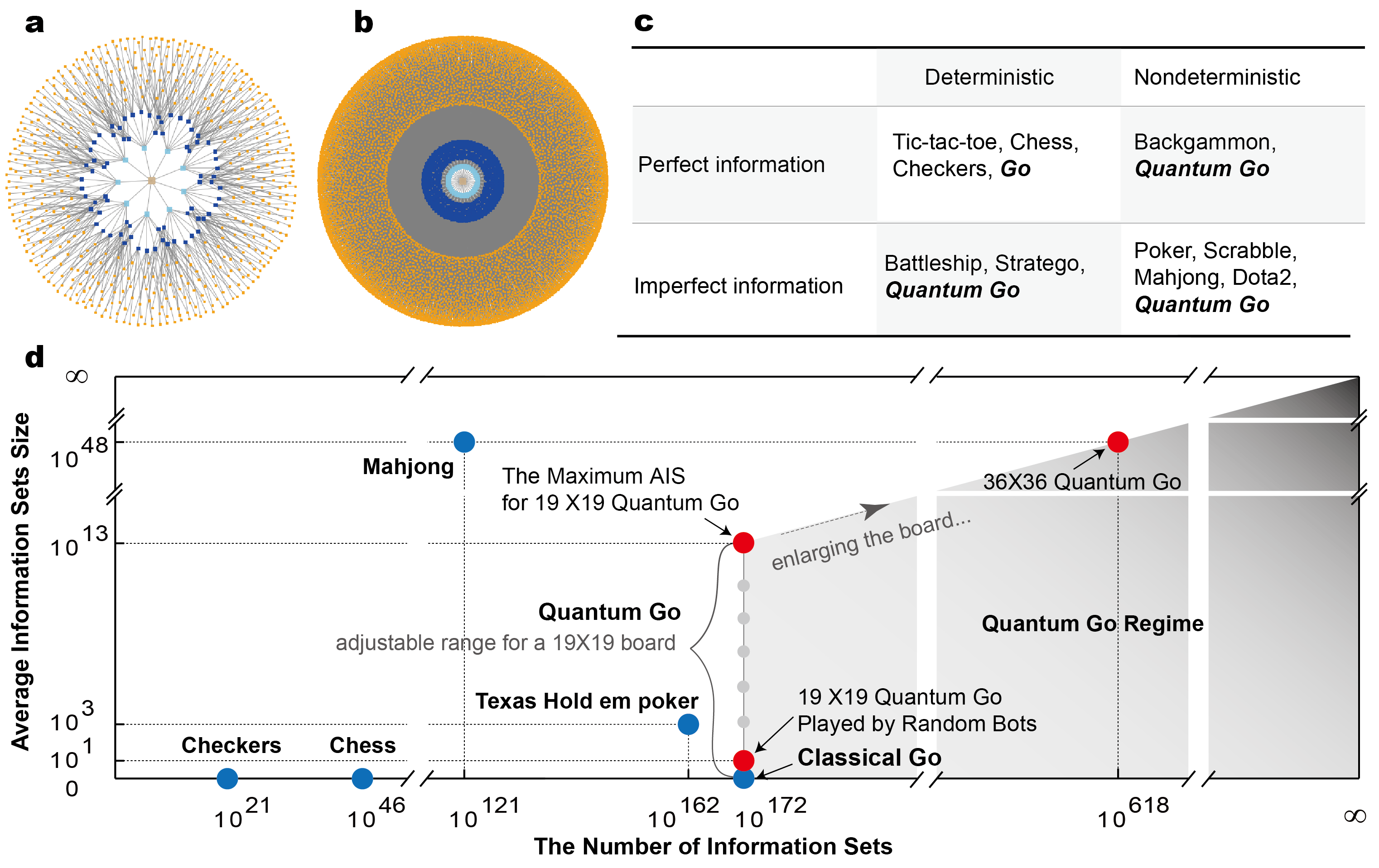}\\
 \caption{\textbf{The complexity of games.} \textbf{a-b}, Comparisons of the game tree sizes between the classical and quantum \emph{Go} on a $3\times3$ board. \textbf{c}, The classification of games based on their features: deterministic/nondeterministic and perfect/imperfect information. \textbf{d}, A comparison of different games\cite{Zha2019} in their (average information sets size) AIS and the number of information sets. The number of information sets represents the information of all observable states while AIS represents the size of hidden information. The game of \emph{Go} has the largest size of state space in classical games, however, it is still a perfect information game. Poker and Mahjong as imperfect information games are considered more difficult for artificial intelligence to solve. The maximum AIS of the quantum \emph{Go} played on a $N \times N$ board can be calculated (more details in the discussion section): maximum AIS = $2^{\lfloor N \times N/8 \rfloor}$. The number of information sets of quantum \emph{Go} is the same as classical \emph{Go}, which is $3^{N \times N}$, since every intersection has three possible states (occupied by white stone, occupied by black stone, and empty). We experimentally demonstrate that quantum \emph{Go} has a wide range of AIS which can be adjusted from 0 to $10^{13}$ on a $19\times19$ board. Especially, we develop a computer program to play the quantum \emph{Go} randomly, and the statistical data shows that the AIS approaches to $10^1$ (see Methods). The AIS of the quantum \emph{Go} played on a $36 \times 36$ board can be up to $10^{48}$, which equals to the AIS of Mahjong. In principle, as we can increase the board size infinitely, quantum \emph{Go} can simulate all games in terms of difficulties in the quantum \emph{Go} regime as shown in the picture. }
  \label{Figure5}
\end{figure*}

\subsection{A demo of quantum \emph{Go} Kifu}

Fig.\ref{Figure4} illustrates a Kifu for two players, Alice and Bob, on a $7\times7$ quantum \emph{Go} board. Alice plays black stones and Bob plays white stones. At first, Alice puts \stoneblack{1} on the intersections F2 and F5. She sets F2 as $p_1$ and F5 as $p_2$ as defined in Eq.(\ref{Equantion1}) in the quantum Go board, which are kept secret to Bob. Then, Bob puts \stonewhite{2} on B6 (as $p_1$) and B2 (as $p_2$), and the choice of $p_1$ and $p_2$ is also unknown to Alice. The game proceeds until Alice puts \stoneblack{5} at D2 (as $p_1$) and B1 (as $p_2$). As B1 is adjacent to B2, \stonewhite{2} and \stoneblack{5} are within each others detectable area, leading to the collapse measurement. The measuring order can be specified as\cite{Ranchin2016}: (1), The stones in the same color with the last stone being placed (not including the last stone) on the board first collapse.
(2), The stones in the other color proceed to collapse measurement.
(3), The last stone proceeds to the final collapse measurement.

So, \stonewhite{2} is the first stone to be measured. We illustrate this game by using the time series data displayed in Fig.\ref{Figure3}\textbf{a} starting from 0.0295s. As shown in both Fig.\ref{Figure3}\textbf{a} and Fig.\ref{Figure4}\textbf{b}, the first measurement result of \stonewhite{2} is ``0''. The result of ``0'' (``1'') corresponding to the quantum stone collapses to $p_1$ ($p_2$). Since Bob sets B6 as $p_1$, \stonewhite{2} collapses to B6. Then Alice's \stoneblack{5} is measured, the result ``0'' tells the stone to collapse to D2. After the collapse measurement, the game returns to a normal turn which begins with Bob placing \stonewhite{6}. The game continues, Fig.\ref{Figure4}\textbf{a} shows the states of the board when the game should step into the collapse measurement process. 

For the moment when Alice places \stoneblacksmall{17} to occupy B2 and B4, \stonewhitesmall{16} becomes surrounded by four black stones. In classical \emph{Go}, \stonewhitesmall{16} will be captured in this situation. Since quantum stones will not occupy neighbors' liberties, \stonewhitesmall{16} will survive instead. \stoneblacksmall{17} is then measured to be a classical stone and collapses to B4, leaving B2 empty again. Since Bob becomes aware of the threat of capturing stone \stonewhitesmall{16}, he puts \stonewhitesmall{18} on B2 as one of the superposition locations. However, Bob fails to occupy B2 after \stonewhitesmall{18} collapses to a classical stone. As \stoneblacksmall{19} is put on the intersections of B2 and F4, the measurement result ``0'' leads to the stone collapse to position $p_1$, which is B2. As a result, \stoneblacksmall{19} occupies B2 as a classical stone which reduces the only one liberty remained for \stonewhitesmall{16} and conducts a capture successfully. Apparently, there exists uncertainty in the defense and attack in quantum \emph{Go} due to its nondeterministic feature. The states of the game do not only depend on the players' choices but also the inherent randomness endowed by the nature.

For the ultimate scenario where the state $\arrowvert{\psi}\rangle_{\tau_{n}}$ is tuned to be very biased down to $\arrowvert{H}\rangle\arrowvert{V}\rangle$, the measurement result will all approach to ``0'' in the time series data. In this situation, all the game states are determined by players' choices so that quantum \emph{Go} becomes a deterministic game. The players keep their own private information before all their quantum stones are measured so that quantum \emph{Go} becomes an imperfect information game. For example, as shown in Fig.\ref{Figure4}, Bob puts \stonewhite{2} at B6 and B2, and set B6 as $p_1$. He can predict that \stonewhite{2} will collapse to B6 after the measurement, but Alice has no idea about that. In this way, Bob can put \stonewhitesmall{18} at B2 as $p_1$ that will protect \stonewhitesmall{16} with a high success possibility.

\section*{Discussion}

With the above game rules and experimental demonstration, it is intuitively to ask whether the quantum version of \emph{Go} surpasses the classical \emph{Go} in the game complexity, while the complexity of classical \emph{Go} is considered to be EXPTIME-complete\cite{Robson1983}. We build a toy model to compare the game tree size between the classical and quantum versions of \emph{Go}. Due to the superposition principle of quantum mechanics, the possible moves of quantum \emph{Go} scale as $S=\binom{N}{2}$, where $N$ is the number of unoccupied intersections left on the board. Fig.\ref{Figure5}\textbf{a-b} illustrate the game tree size of classical and quantum \emph{Go} for a $3\times3$ board within three steps. We can see that quantum \emph{Go} is much more complex than its classical counterpart. With the board size increasing, the complexity of quantum \emph{Go} will grow much faster than classical \emph{Go}, and become impossible to visualize properly. The rigorous relation between the complexity and the difficulty of quantum \emph{Go} is still an open question, which may inspire new research topics for mathematics and computer sciences.

Quantum \emph{Go} is different from all the other nondeterministic and imperfect information games since it can cover a wide range of game difficulties rather than a single point, which gives a wide range of benchmarks for artificial intelligence. For comparison, we list quantum \emph{Go} with a few typical games with classification in Fig.\ref{Figure5}\textbf{c}. As a perfect information game, \emph{Go} has the largest number of game states among all classical games. In imperfect information games, average information sets size (AIS) is used to quantify the private information in the game\cite{Johanson2013}. The game with large AIS is believed to be more challenging than the perfect information game. The AIS of quantum \emph{Go} is changeable by tuning the entangled states. For the ultimate scenario of $\arrowvert{H}\rangle\arrowvert{V}\rangle$, the collapse result depends on the player's choice solely, and the quantum \emph{Go} has the largest AIS. In a $19\times19$ board, there are 361 intersections to fit a maximum number of pairs up to 90, supposing no collapse. Since each player owns 45 pairs of states, the AIS can be up to $2^{45} \approx 10^{13}$. More generally, the maximum AIS can be calculated by $2^{\lfloor N \times N/8 \rfloor}$, where $N\ge2$. If the state is maximally entangled, then the collapse results all depend on the quantum mechanics with inherent randomness. In this case, both players have no private information in this situation, and their AIS is 0. In the cases where non-maximally entangled states are adopted, both players' choices and quantum randomness will have an impact on the outcome, resulting in the values of AIS ranging from 0 to $10^{13}$ as shown in Fig.\ref{Figure5}\textbf{d}. We run the quantum \emph{Go} games by two bots who choose the intersections to put their stones randomly. The statistical results show that the AIS approach to $10^{1}$ by stochastic playing (see Methods). In principle, as we can increase the board size infinitely, quantum \emph{Go} can simulate all games in terms of difficulties in the quantum \emph{Go} regime.\\

\section*{Conclusion}

In conclusion, we have proposed and demonstrated the quantum version of the ancient game \emph{Go} with entangled photon pairs and time series scheme. 
Experimentally obtained terabyte of time series data generate a huge dimension of Hilbert space, and the series obtained in one hour may support a game with the AIS up to $2^{10^8}$ and the moves up to $10^8$. We have also investigated quantum \emph{Go} in the regime of nondeterministic and imperfect information games by tuning the quantum stones from maximally entangled states to non-maximally entangled states until to product states. Comparing a few typical games, we have found that quantum \emph{Go} can cover a wide range of game difficulties rather than a single point, suggesting a versatile and promising platform for testing new algorithms for artificial intelligence.

\section*{Acknowledgments}
The authors thank Jian-Wei Pan for helpful discussions. This research was supported by the National Key R\&D Program of China (2019YFA0308700, 2017YFA0303700), the National Natural Science Foundation of China (61734005, 11761141014, 11690033), the Science and Technology Commission of Shanghai Municipality (STCSM) (17JC1400403), and the Shanghai Municipal Education Commission (SMEC) (2017-01-07-00-02- E00049). X.-M.J. acknowledges additional support from a Shanghai talent program.\\

\section*{Methods}

 \noindent\textbf{Rule of Classical \emph{Go} and the comparison with Quantum \emph{Go}.} Classical \emph{Go} is a game played on a board. The standard \emph{Go} board is a square grid of 19 horizontal and 19 vertical lines, containing 361 intersections. The playing pieces are called stones that can be placed on the intersections. Classical \emph{Go} can also be played on the board of other sizes, for example, $9\times9$ and $13 \times 13$ are common sizes for beginners or for  those who want quick games.

The basic rules of classical \emph{Go} include:

 \noindent\uppercase\expandafter{\romannumeral1}. It is played by two players.

 \noindent\uppercase\expandafter{\romannumeral2}. Two players place their stones on the board alternately with one using black stones while the other using white ones.

 \noindent\uppercase\expandafter{\romannumeral3}.  In each turn, each player can place one stone on one intersection of the board, but they can always choose to do nothing and pass their turn.

 \noindent\uppercase\expandafter{\romannumeral4}. The stones cannot be moved once they are placed on certain intersections, but they can be captured and removed.

\noindent\uppercase\expandafter{\romannumeral5}. The game ends when both players pass their turn consecutively.

The stones will be captured when completely surrounded by the other players' classical stones. We call direct adjacent (up, down, left, or right) stones are the neighbor stones. The neighbor stones with the same color form a connected group of stones. In classical \emph{Go}, an empty intersection gives a liberty to the neighbor stone or the neighbor stones group. To stay on the board, a group needs one liberty at least. Filling the opponent's liberties is the way to capture his/her stones. As shown in Extended Data Fig.1\textbf{a-b}, three black stones are connected, which share two liberties as a group. If there are two classical white stones placed on the intersections marked with squares (Extended Data Fig.1\textbf{a}), these black stones are captured as no liberty is left. While in quantum \emph{Go}, besides empty intersections, the intersection occupied by quantum stones also gives their neighbors liberties (Extended Data Fig.1\textbf{c}). After the collapse measurement, the quantum stones will become classical stones. 

In quantum \emph{Go}, if the detectable area of a quantum stone contains other stones, the quantum stone will be measured. In this work, we set direct adjacent intersections as the detectable areas (Extended Data Fig.1\textbf{d}), while the detectable areas can be set to bigger or smaller range which will make the game features different.

The player possesses a larger section of the board will win the game. The ending condition and scoring system of quantum \emph{Go} is similar to classical \emph{Go}.\\

 \noindent\textbf{Quantum \emph{Go} as an imperfect information game.} Two players take turns to place their quantum stones. They have to make their strategies at the presence of other quantum stones, the positions of which are undetermined until being measured. The existence of quantum stones makes quantum \emph{Go} an imperfect information game, which makes it difficult for players to make strategies. Other than counting the number of game states, we count the number of information sets (which is equal to the decision points of the game) in imperfect information games\cite{Johanson2013}. An information set of one player is a collection of game states among which he/she cannot distinguish.

In the game state of Move 3 shown in Fig.1\textbf{a}, there are 2 black quantum stones on board which can result in four possible game states after the measurement (Fig.1\textbf{b}). The white player has no idea which state of these black quantum stones will collapse to, so they are all in one information set which are indistinguishable for the white player to make strategy when placing \stonewhite{4}.  For example, the first move can backtrack, which the black player place \stoneblack{1} on C5 and C3. The black player sets C5 as $p_1$ and C3 as $p_2$ which is recorded in Fig.1\textbf{c}, but it is kept secret to the white player. The state of \stoneblack{1} can be written as $a_1\arrowvert{1}\rangle_{C5}\arrowvert{0}\rangle_{C3}+a_2\arrowvert{0}\rangle_{C5}\arrowvert{1}\rangle_{C3}$. While in the view of the white player, the state can be either $a_1\arrowvert{1}\rangle_{C5}\arrowvert{0}\rangle_{C3}+a_2\arrowvert{0}\rangle_{C5}\arrowvert{1}\rangle_{C3}$ or $a_1\arrowvert{1}\rangle_{C3}\arrowvert{0}\rangle_{C5}+a_2\arrowvert{0}\rangle_{C3}\arrowvert{1}\rangle_{C5}$.

In each move of the game, $n$ quantum stones of one player can be represented by a complex vector in the Hilbert space of $2^n$ dimensions, with $2^n$ possible classical states in the other's information sets.\\

 \noindent\textbf{Quantum stone box.} In classical \emph{Go}, each player has a stone box with different colors, one in black and the other in white. In quantum \emph{Go}, both players can put quantum stones obtained from one box and get their quantum stone states stored in the time series data. All the quantum stone states are identical until they collapse. In this work, we use polarization entangled photon pairs as the quantum stones for both players to demonstrate the scheme. Here we define $\arrowvert{H}\rangle=\arrowvert{1}\rangle$ and $\arrowvert{V}\rangle=\arrowvert{0}\rangle$. As shown in Fig.2\textbf{a}, we generate the polarization entangled photon pairs through type-II spontaneous parametric down conversion\cite{Kim1996}. We use the quasi-phase-matched periodically-poled $KTiOPO_4$ (PPKTP) crystal and the crystal is bi-directionally pumped in a Sagnac interferometer. The 405nm pump laser first passes through a combination of a polarizing beamsplitter(PBS), a half wave plate(HWP) and quarter wave plate(QWP). A superposition state of the pump laser $\cos\theta\arrowvert{1}\rangle+e^{i\phi}\sin\theta\arrowvert{0}\rangle$ can be prepared by the combination of HWP and QWP. The pump laser passes through a dichroic mirror, which transmits the ultraviolet light and reflects the infrared light. The pump laser is guided into a Sagnac-loop which consists of a PPKTP crystal, a dual-wavelength PBS and a dual-wavelength HWP (set at 45 degree). The PBS divides the pump laser into two directions (the clockwise and the counterclockwise) and are all focused into the PPKTP crystal. The down-converted photons generated by the two different directions interfere at the PBS and become indistinguishable through careful alignment of the Sagnac interferometer. The interferometer generates the following entangled state,
\begin{equation}
\arrowvert{\psi}\rangle=\cos\theta\arrowvert{1}\rangle\arrowvert{0}\rangle+e^{i\phi}\sin\theta\arrowvert{0}\rangle\arrowvert{1}\rangle.
\end{equation}
By tuning the parameters $\theta$ and $\phi$, we can engineer the state into different quantum stones to investigate quantum \emph{Go} in the regime of nondeterministic and imperfect information games. When $\theta=45^\circ$, $\phi=0^\circ$, we get the maximally entangled states, and when $\theta=0^\circ$, $\phi=90^\circ$, the state becomes separable.\\

 \noindent\textbf{Time-of-flight record and storage of quantum stones.} 
In this experiment, we measure the single-photon events with high time precision. As shown in Fig.2\textbf{b}, after the polarization information is measured by the PBSs, the photons are directed into four single photon detectors respectively. The four output channels connect the FPGA(Field Programmable Gate Array) in the time-of-flight storage module(Fig.2\textbf{c}).

Time-of-flight is a method usually used to measure the distance between an object and a sensor, by recording and calculating the time difference between different light paths\cite{Delpy1988}. In this work, we use the time-of-flight technique at single photon level. The tremendous data of arrival time of each photon will be transmitted and recorded, which used to be considered as an intractable task. With high-performance FPGA, high-speed digital transmission technique and processing software, the time-of-flight record of signal photons in different channels can be stored as time-labelled data. The coincidence events of correlated photon pairs can be extracted by setting the coincidence window and proper time delay of different channels.

The coincidence events are encoded into a 0/1 sequence which has the inherent randomness, because the measurement result of one entangled pair can not be speculated by the results of all the other entangled pairs produced by the same photon source. To demonstrate it, we calculate the autocorrelation function of the time series data of different entangled states (Extended Data Fig.2).\\

 \noindent\textbf{The autocorrelation function of time series.}  Autocorrelation, also called as serial correlation or lagged correlation, is the correlation of a series with a lagged copy of itself. If a time series is autocorrelated, the series is predictable as the futures value has a  relation with the past values. Many physical time series are autocorrelated, as inertia in the physical system makes the past states affect the present state. However, the quantum time series is an exception.

If there are two time series $x$ and $y$ with length $N$, the correlation coefficient is given by
$r=\frac{\sum_{t=1}^{N}(x_{t}-\bar{x})(y_{t}-\bar{y})}{{[\sum_{t=1}^{N}{(x_{t}-\bar{x})}^{2}]}^{\frac{1}{2}}{[\sum_{1}^{N}{(y_{t}-\bar{y})}^{2}]}^{\frac{1}{2}}}$. The correlation coefficient of the successive observations in one time series is similar, which is computed between the series and its lagged copy by $k$ units($1<k<N$, $k\in\mathbb{N}$). The autocorrelation function is a function of time lag and the autocorrelation coefficient at lag $k$ can be expressed as $r_{k}=\frac{\sum_{t=1}^{N-k}(x_{t}-\bar{x}_{(1)})(x_{t+k}-\bar{x}_{(2)})}{{[\sum_{t=1}^{N-k}{(x_{t}-\bar{x}_{(1)})}^{2}]}^{\frac{1}{2}}{[\sum_{t=k+1}^{N}{(x_{t}-\bar{x}_{(2)})}^{2}]}^{\frac{1}{2}}}$, where $\bar{x}_{(1)}$ represents the mean of first $N - k$ observations and $\bar{x}_{(2)}$ is the mean of last $N - k$ observations.
If $N$ is reasonably large and $N \gg k$, $r_{k}$ can be approximated by
$r=\frac{\sum_{t=1}^{N-k}(x_{t}-\bar{x})(x_{t+k}-\bar{x})}{{\sum_{t=1}^{N}{(x_{t}-\bar{x})}^{2}}}$, where $\bar{x}$ is the mean of the overall observations.

If a time series $x$ is random, the lagged autocorrelation coefficients $r_{k}$ are normally distributed with a mean value of 0 and a variance of $1/N$, where $N$ is the sample size. The $95\%$ confidence limits approximate $0 \pm\frac{2}{\sqrt{N}}$.

We calculate lagged autocorrelation coefficients of the time series obtained from different entangled states. The size of the test sample is $N = 200000$ and the lags are from 1 to 10000. As shown Extended Data Fig.2\textbf{a}, two red lines are the $95\%$ confidence limits, which are calculated to be  $\pm\frac{2}{\sqrt{200000}}= \pm0.0045$. There are 10000 data points in the diagram and 490 data points are outside the two red lines, which has a proportion of $4.9\%$. The data distribution is shown in the right insert graph,  the test data outside the $95\%$ confidence limits is all less than $5\%$, which means that the time series is not autocorrelated.\\
 
 \noindent\textbf{Stochastically  playing quantum \emph{Go}.} The human-played Kifu is lacking for quantum \emph{Go}. Here we develop bots to play the game, which can produce a large number of Kifu in a short time. In this work, the bots are two naive bots who stochastically place the stones, which is enough to get the statistical data of games. In the future work, the bots will evolve by self-playing using reinforcement learning. 

The number of quantum stones on the board determines the size of the information sets of each move. The AIS (average information sets size) $S_{avg.infoset}$, which is a metric for imperfect information games, can be calculated by the average number of quantum stones on board. The bigger the AIS is, the more imperfect information the game has. As the number of legal moves in each game is uncertain, we use $S_{avg.infoset}^{N}$ to represent the AIS when the game ends with N moves.

Extended Data Fig.3\textbf{a} illustrates the number of quantum stones in 10 games. 
$Q_i$ is the number of quantum stones on the board in move $i$. The maximum $Q_i$ indicates that in general there are at most tens of quantum stones on the board for a $19\times19$ stochastic game. While almost all $Q_i=0$ when $i > 180$, which indicates the quantum stones will be measured right after being placed in the late game, as there are not many free intersections left. 

As each player's information sets are decided by the other player's quantum stones, we count the number of quantum stones in different colors before calculating the information sets size.
The black(white) player places the stones only in the odd(even) number moves, so we calculate the average number of black(white) quantum stones at odd(even) number moves respectively.
The average number of the white quantum stones at move $N$ (when $N$ is an even number) is $Q_{avg}^N=(\sum_{i=0}^{\frac{N}{2}}Q_{2i})/(\frac{N}{2}+1)$.
The average number of the black quantum stones at move $N$ (when $N$ is an odd number) is $Q_{avg}^N=(\sum_{i=1}^{\frac{N+1}{2}}Q_{2i-1})/(\frac{N+1}{2})$.
The average information sets size is: $S_{avg.infoset}^{N}={\binom{2}{1}}^{Q_{avg}^{N-1}}$.

Extended Data Fig.3\textbf{b-c} show the average number of the white and black quantum stones. The insert figures give the statistical values for 150 games, the blue dots are the mean value and the gray bars are the standard deviations. The average number of the black quantum stone is slightly larger than the white quantum stone (Extended Data Fig.3\textbf{d}), which leads to the average information sets size of the white player slightly larger than the black player (Extended Data Fig.3\textbf{e}). It means in quantum \emph{Go}, the black player not only has the advantage of moving first, but also has the advantage of making strategies. The game can be balanced by setting the proper compensation points which is called komi in classical \emph{Go}. The result of stochastic moves shows AIS can reach $10^1$ for a $19\times19$ board.

{}

\begin{figure*}
\centering
\includegraphics[width=1.98\columnwidth]{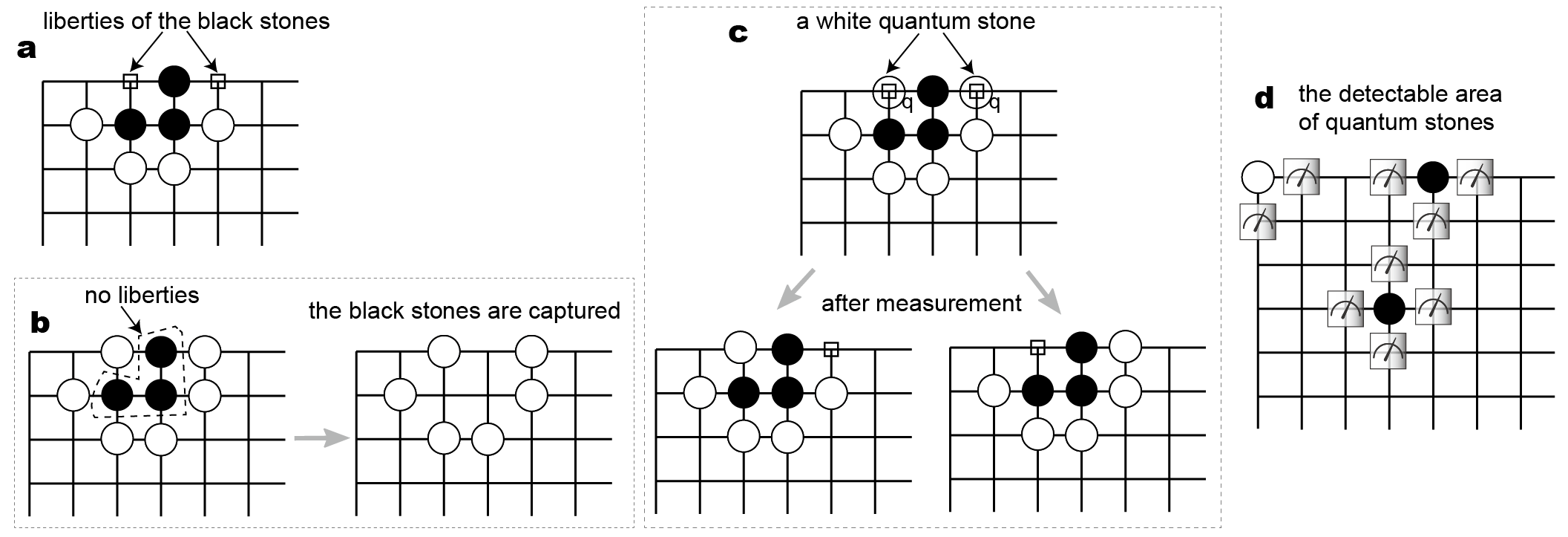}\\
\flushleft
Extended Data Fig. 1. \textbf{The liberties and detectable area.} \textbf{a}, The three black stones are connected together, which has two liberties on the intersections marked with squares. \textbf{b}, If a stone or the group of connected stones has no liberty, they will be captured. \textbf{c}, one may intend to place a quantum stone to capture the other's stones. But it won't work as quantum stone will not occupy the liberties of neighbors' stones. Once the quantum stone is placed on intersections that have neighbors, it will be measured and collapse to be a classical stone. After the measurement, there are two possible classical game states. \textbf{d}, In this work, the detectable area of a quantum stone is the direct adjacent (up, down, left and right) intersections.
\label{ExtendedDataFig1}
\end{figure*}

\begin{figure*}
 \centering
 \includegraphics[width=1.98\columnwidth]{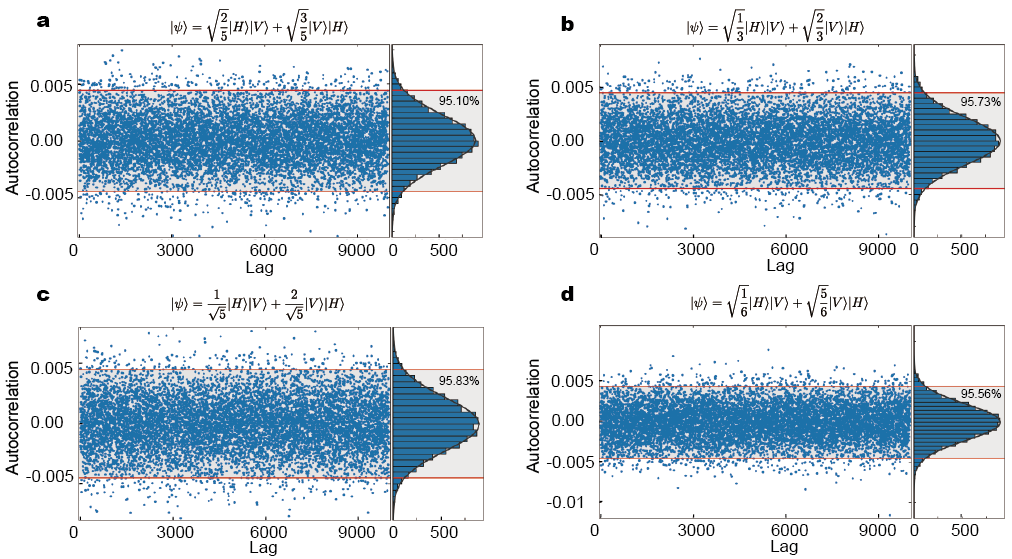}\\
\flushleft
Extended Data Fig. 2.\textbf{The correlograms for 4 different entangled states.} \textbf{a}, The proportion of ``0'' and ``1'' are expected to be $2:3 $ in the time series generated from the entangled state $\arrowvert{\psi}\rangle=\sqrt{\frac{2}{5}}\arrowvert{H}\rangle\arrowvert{V}\rangle+\sqrt{\frac{3}{5}}\arrowvert{V}\rangle\arrowvert{H}\rangle$. While successive observations of the time series are also not correlated. The left graph is the scatterplot with lagged  autocorrelation coefficients. The two red lines give the $95\%$ confidence limits for the series that being not autocorrelated. The histograms on the right give the distribution of these coefficients. It is shown that $95.10\%$ coefficients are inside two red lines which is beyond the $95\%$ confidence limits. \textbf{b-d}, The correlograms of the other three entangled states. The time series are all not autocorrelated with high confidence.
 \label{Extended Data Fig. 2}
\end{figure*}

\begin{figure*}
 \centering
 \includegraphics[width=1.98\columnwidth]{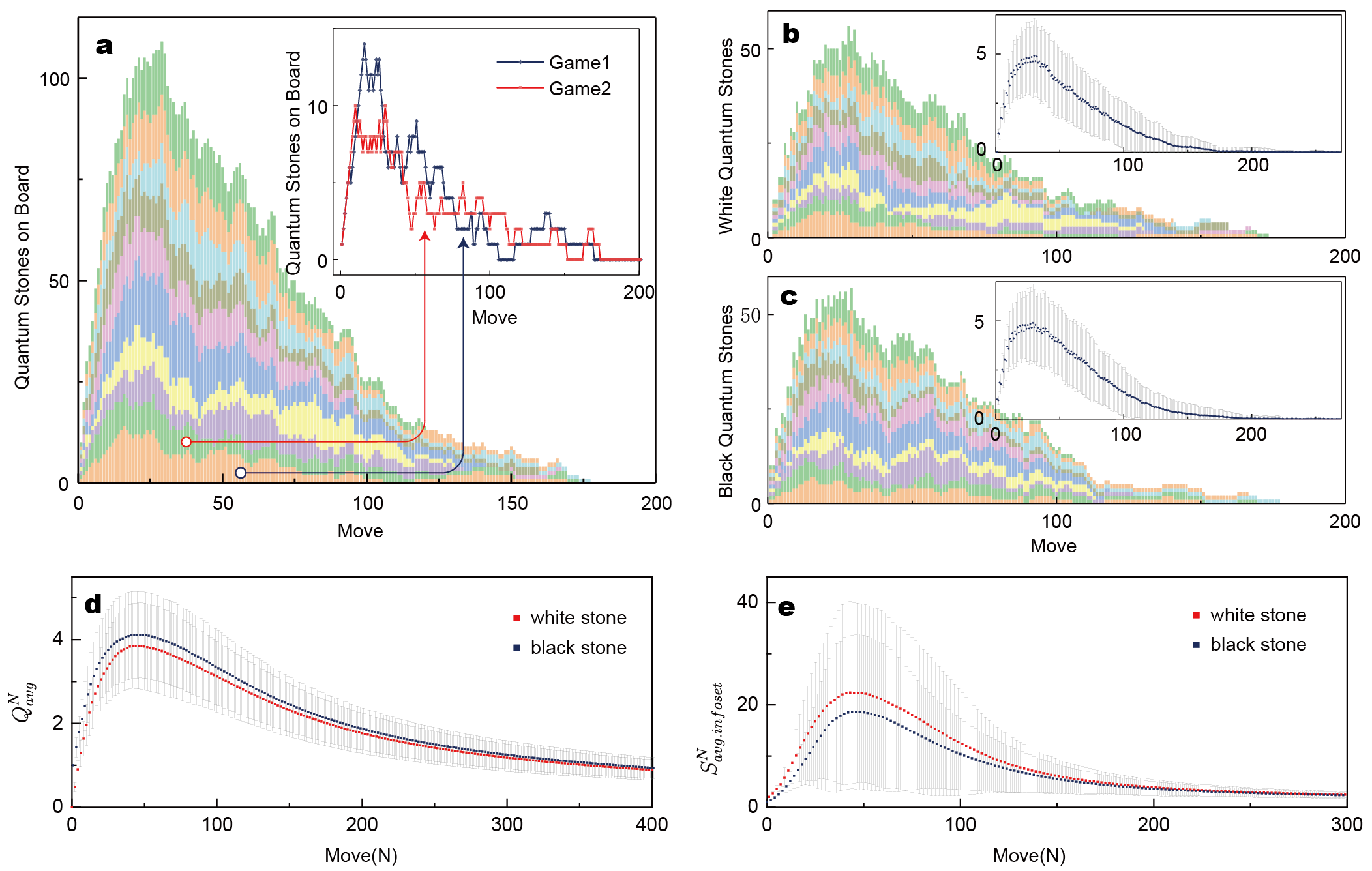}\\
\flushleft
Extended Data Fig. 3. \textbf{Statistical data of stochastically playing.} \textbf{a}, Ten-games samples of quantum \emph{Go}. In each move, the number of quantum stones is counted. After 175 moves, there are almost no quantum stones on board. \textbf{b-c}, the number of white/black quantum stones on board of each move. The insert graph shows the statistical result (blue dots are the mean values and the grey error bars are the standard deviations).  \textbf{d}, the average number of quantum stones $Q_{avg}^N$ on board. \textbf{e}, the average information set scale $S_{avg.infoset}^{N}$ is also a function of move $N$. The $S_{avg.infoset}^{N}$ for black(white) is calculated at odd(even) moves, as black(white) only plays at odd(even) moves.
\label{Extended Data Fig. 3}
\end{figure*}

\clearpage

\section*{Supplementary Materials: Quantum \emph{Go} Machine}

\section*{A Complete Kifu}

Some readers may be interested in the game of quantum \emph{Go}, and want to play it with a quick start. So we provide a complete Kifu in Kifu.dat(Can be opened with any text editor). Hundreds of moves in the Kifu will make readers familiar with the rules of the game.

In addition to the game states, the number of stones and the average information sets size are also given in each move. The average number of white quantum stones at move N (when N is an even number) is $Q_{avg}^N=(\sum_{i=0}^{\frac{N}{2}}Q_{2i})/(\frac{N}{2}+1)$.
The average number of black quantum stones at move N (when N is an odd number) is $Q_{avg}^N=(\sum_{i=1}^{\frac{N+1}{2}}Q_{2i-1})/(\frac{N+1}{2})$.
The average information sets size is: $S_{avg.infoset}^{N}={\binom{2}{1}}^{Q_{avg}^{N-1}}$. In the following, We use some game states to illustrate how to calculate these parameters by using these formulas. 

The first four moves of a game played by the bots are shown in Fig.S1.
There is no stone on the board at the start ($Q_{avg}^{0}=0$), so $S_{avg.infoset}^1={\binom{2}{1}}^{0}=1$ for the black player (as the black player first to play).
After move 1, there is one black quantum stone on the board ($Q_{avg}^{1}=1$), so $S_{avg.infoset}^2=2$ for the white player who is next to play. 
 After move 2, there is one white quantum stone on board. So the average number of white quantum stones on board is $(0+1)/2=0.5$, and $S_{avg.infoset}^3=1.4$ for the black player in move 3. In the same way, we can calculate that $S_{avg.infoset}^4=2.8$ and $S_{avg.infoset}^5=2$.

In each turn, one quantum stone will be added to the board if no player passes, while the number of quantum stones will reduce when the collapse measurement takes place. In Fig.S2\textbf{a}, white10 is placed on C16 and L10 which causes the collapse measurement. After the measurement, there remains 4 white quantum stones on board, so $Q_{avg}^{10}=(\sum_{i=0}^{5}Q_{2i})/(5+1)=(4+\sum_{n=0}^{4}n)/6=\frac{7}{3}$ and $S_{avg.infoset}^{11}={\binom{2}{1}}^{\frac{7}{3}}=5.04$. 

In quantum \emph{Go}, the rule of capturing stones is similar to classical \emph{Go}. In Fig.S2\textbf{c-d}, the white classical stone on A1 is captured after the black quantum stone [B1,B10] becoming a classical stone that settled on B1. The self-capture rule and the Ko rule are also included in quantum \emph{Go}, which is the same as classical \emph{Go}.

As the games played by the naive bots,  the boards of final states are almost filled with no legal intersections remaining to place the stones. The games ended as two bots pass the turns successively. In hundreds runs of games, the bots end the games in 400-600 moves.  Fig.S3 shows the final board state of one game. The winner is the black player, which has a winning margin with 50 points, when using area scoring and komi=0. Usually, komi will be set as 6.5 or 7.5 in classical \emph{Go}, since the black player has an advantage to place stone first. But in the game of stochastically playing, there is little advantage for the black player. In 150 stochastic games, black wins 76 games when komi is 0.

{}

\begin{figure*}
  \centering
 \includegraphics[width=1.98\columnwidth]{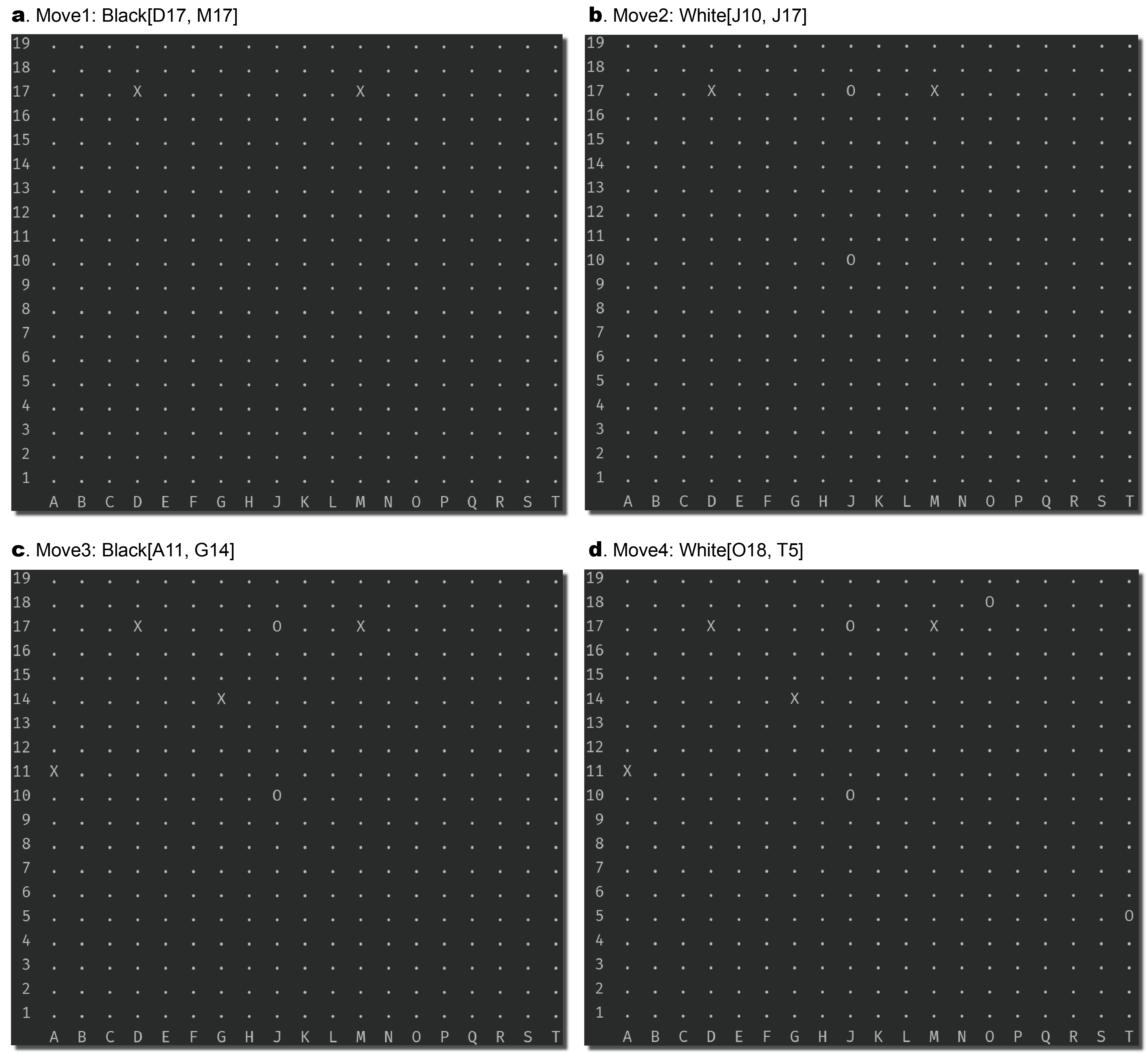}\\
\flushleft
 Figure S1:
   \textbf{The average information sets size for the first four moves of a game.} Upper case X represents black quantum stone while O represents white quantum stone. Lower case x represents black classical stone while o represents white classical stone. Dots represent the empty intersections. \textbf{a}, In move 1, after the black player places a quantum stone on D17 and M17, there is one black quantum stone on the board($Q_{avg}^{1}=1$), so $S_{avg.infoset}^2={\binom{2}{1}}^{1}=2$ for the white player who is next to play. 
\textbf{b}, After the white player places a quantum stone on J10 and J17 in move 2, there are two quantum stones (one black and one white) on the board. So the average number of white quantum stones on board is $(\sum_{n=0}^{\frac{2}{2}}Q_{2i})/(\frac{2}{2}+1)=(0+1)/2=0.5$, and $S_{avg.infoset}^3={\binom{2}{1}}^{0.5}=1.4$ for the black player in move 3. 
\textbf{c}, There are 2 black quantum stones on the board after black3 is placed, the average number of black quantum stones on board is $(\sum_{i=1}^{\frac{3+1}{2}}Q_{2i-1})/(\frac{3+1}{2})=(1+2)/2=1.5$. The information set size for the white player in move 4 is $S_{avg.infoset}^4={\binom{2}{1}}^{1.5}=2.8$. \textbf{d}, In the same way, $S_{avg.infoset}^5={\binom{2}{1}}^{1}=2$ for the black player. 
  \label{FigureS1}
\end{figure*}

\begin{figure*}
  \centering
 \includegraphics[width=1.98\columnwidth]{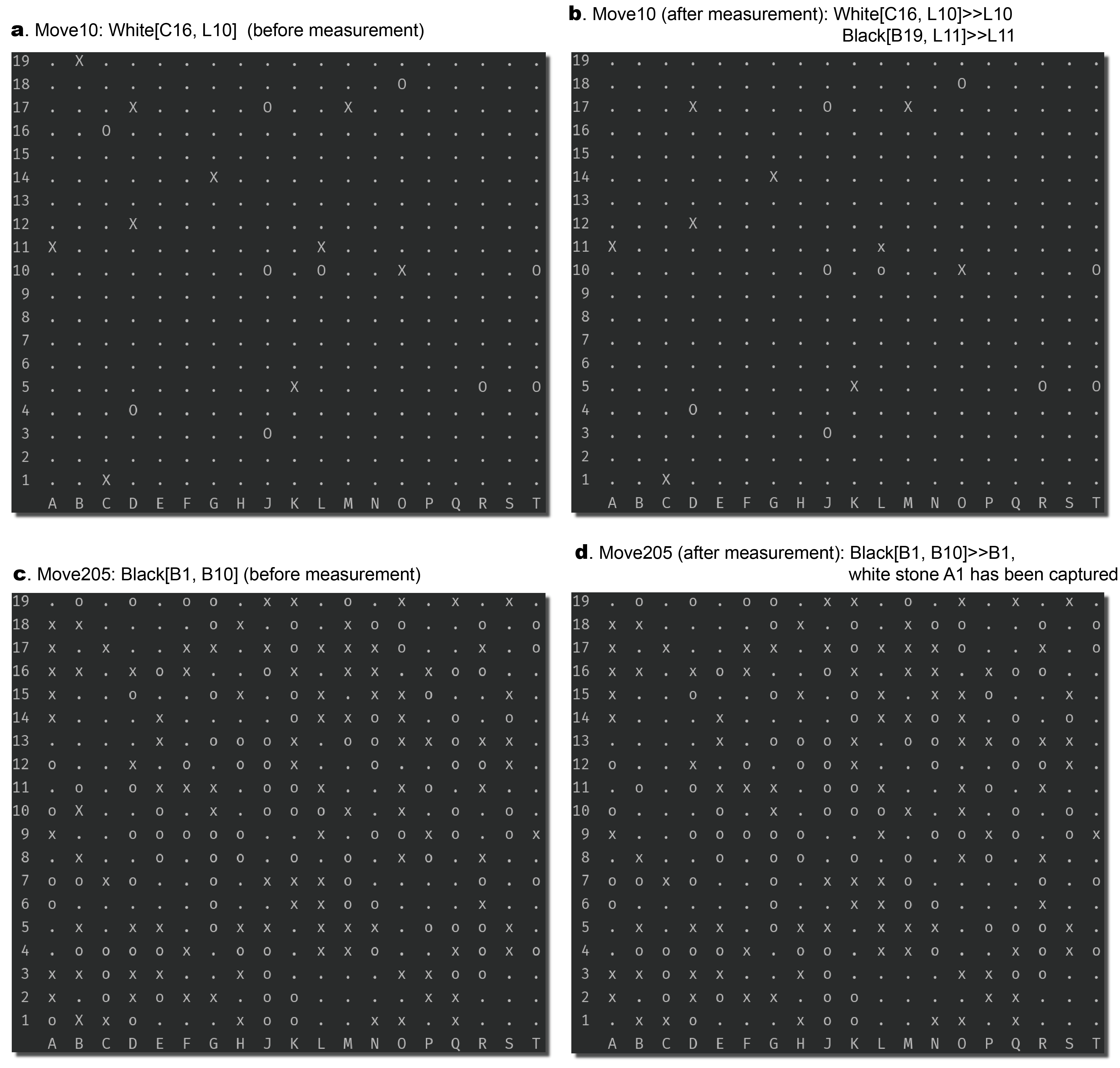}\\
 \flushleft
 Figure S2:
   \textbf{Collapse measurement and stone capture.} \textbf{a-b}, In Move 10, the white quantum stone is placed on C16 and L10, which causes the collapse measurement. Two stones are measured, the white quantum stone(on [C16, L10]) is collapse to L10, and the black quantum stone(on [B19, L11]) is collapse to L11. After the measurement, there remains 4 black quantum stones and 4 white stones on the board. The average number of  white quantum stones until this move is $Q_{avg}^{10}=(\sum_{i=0}^{5}Q_{2i})/(5+1)$$=(4+\sum_{n=0}^{4}n)/6=\frac{7}{3}$, and $S_{avg.infoset}^{11}={\binom{2}{1}}^{\frac{7}{3}}=5.04$.  \textbf{c-d}, The black quantum stone is placed on B1 and B10 and collapses to B1 as a classical stone after the measurement. It fills up the liberties of the classical white stone on A1, which makes the stone been captured. 
  \label{FigureS2}
\end{figure*}

\begin{figure*}
  \centering
 \includegraphics[width=1.98\columnwidth]{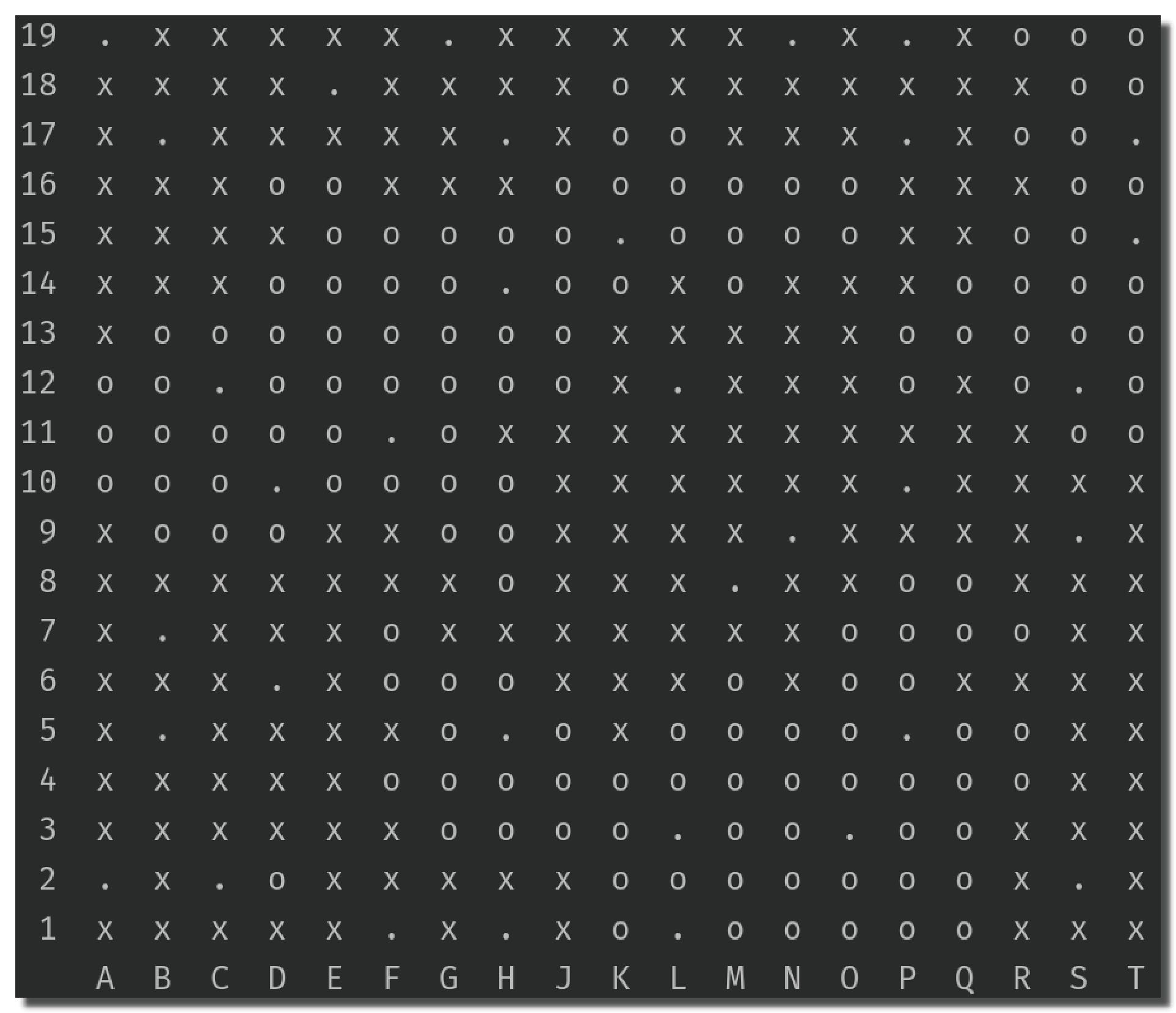}\\
  \flushleft
 Figure S3:
   \textbf{A final board state.}The game ends as the two players pass the turns consecutively. The winner is the black player and winning margin is 50 points when using area scoring and komi = 0. 
  \label{FigureS3}
\end{figure*}


\begin{thebibliography}{}
\bibitem{Schaeffer2000} Schaeffer, J. The games computers (and people) play. \textit{Fortieth Anniversary Volume: Advancing into the 21st Century} \textbf{52}, 189-266 (2000).

\bibitem{Kocsis2006} Kocsis, L. \& Szepesvári, C., Bandit Based Monte-Carlo Planning. \textit{Machine Learning: ECML 2006}, 282-293 (2006).

\bibitem{Coulom2006} Coulom, R. Efficient selectivity and backup operators in Monte-Carlo tree search. In \textit{5th Int. Conf. Computers and Games } 72-83 (2006).

\bibitem{Browne2012} Browne, C. \textit{et al.}  A survey of Monte Carlo tree search methods. \textit{IEEE Trans. Comput. Intell. AI Games} \textbf{4}, 1-49 (2019).

\bibitem{Silver2016} Silver, D. \textit{et al.} Mastering the game of Go with deep neural networks and tree search. \textit{Nature} \textbf{529}, 484-489 (2016).

\bibitem{Silver2017} Silver, D. \textit{et al.} Mastering the game of Go without human knowledge. \textit{Nature} \textbf{550}, 354-359 (2017).

\bibitem{Allis1994}  Allis, L. V. Searching for solutions in games and artificial intelligence. \textit{Ph.D.Thesis, University of Limburg, Maastricht}, (1994).

\bibitem{Pumperla2019} Pumperla, M. \& Ferguson, K. Deep Learning and the Game of Go. (Manning Publications, 2019).

\bibitem{Moravcik2017} BMoravčík, M. \textit{et al.} Deepstack: Expert-level artificial intelligence in heads-up no-limit poker. \textit{Science} \textbf{356}, 508-513 (2018).

\bibitem{Brown2019} Brown, N. \& Sandholm, T. Superhuman AI for multiplayer poker. \textit{Science} \textbf{365}, 885-890 (2019).

\bibitem{Brown2018} Brown, N. \& Sandholm, T. Superhuman AI for heads-up no-limit poker: Libratus beats top professionals. \textit{Science} \textbf{359}, 418-424 (2018).

\bibitem{Vinyals2017} Vinyals, O. \textit{et al.} StarCraft {II:} {A} New Challenge for Reinforcement Learning. \textit{preprint arXiv:}1708.04782 (2017).

\bibitem{Semenov2016} Semenov, A. \textit{et al.} Performance of Machine Learning Algorithms in Predicting Game Outcome from Drafts in Dota 2. In \textit{International Conference on 
Analysis of Images, Social Networks and Texts},  
26-37, (Springer,2016).

\bibitem{Frank1998} Frank, I., Basin, D. \& Matsubara, H.  Finding optimal strategies for imperfect information games. \textit{Proc.
AAAI-98}, 500–507(1998).

\bibitem{Monroe1996} Monroe, C., Meekhof, D. M., King, B.E. \& Wineland, D. J.  A ``Schrodinger Cat'' Superposition State of an Atom. \textit{Science} \textbf{272}, 1131 (1996).

\bibitem{Bennett1984} Bennett, C. H. \&  Brassard, G. \textit{Systems and Signal Processing} 175-179 (1984).

\bibitem{Jin2010}  Jin, X.-M. \textit{et al.} Experimental free-space quantum teleportation. \textit{Nat. Photon.} \textbf{4}, 376 (2010).

\bibitem{Pan2012}  Pan, J.-W. \textit{et al.} Multiphoton entanglement and interferometry. \textit{Rev. Mod. Phys.} \textbf{84}, 777 (2012).

\bibitem{Lo2014}  Lo, H. K., Curty,  M. \& Tamaki, K. Secure quantum key distribution. \textit{Nat. Photon.} \textbf{8}, 595-604 (2014).

\bibitem{Shor1994} Shor, P. W. in \textit{Proceedings of the 35th Annual Symposium on the Foundations of Computer Science} 124-133 (IEEE Computer Society Press, Los Alamitos, California, 1994).

\bibitem{Grover1997} Grover, L. K. Quantum mechanics helps in searching for a needle in a haystack. \textit{Phys. Rev. Lett.} \textbf{79}, 325-328 (1997).

\bibitem{Kocsis2006a} Kocsis, L. \& Szepesvári, C. \textit{Machine Learning: ECML 2006}, 282-293 (2006).

\bibitem{Biamonte2017} Biamonte, J. \textit{et al.} Quantum machine learning. \textit{Nature} \textbf{549}, 195–202 (2017).

\bibitem{Gelly2014} Gelly, S. \textit{et al.} The Grand Challenge of Computer Go: Monte Carlo Tree Search and Extensions. \textit{Commun. ACM} \textbf{55}, 106-113 (2012).

\bibitem{Lloyd2014} Lloyd, S., Mohseni, M. \& Rebentrost, P. Quantum principal component analysis. \textit{Nat. Phys.} \textbf{10}, 631-633 (2014).

\bibitem{Rebentrost2014} Rebentrost, P., Mohseni, M. \& Lloyd, S. Quantum support vector machine for big data classification. \textit{Phys. Rev. Lett.} \textbf{113}, 130503 (2014).

\bibitem{Lloyd2013} Lloyd, S., Mohseni, M. \& Rebentrost, P. Quantum algorithms for supervised and unsupervised machine learning. \textit{preprint arXiv:}1307.0411 (2013).

\bibitem{Melnikov2018} Melnikov, A. A. \textit{et al.} Active learning machine learns to create new quantum experiments. \textit{PNAS} \textbf{115(6)}, 1221-1226 (2018).

\bibitem{Cai2015} Cai, X.-D. \textit{et al.} Entanglement-Based Machine Learning on a Quantum Computer. \textit{Phys. Rev. Lett.} \textbf{114}, 110504 (2015).

\bibitem{Li2015} Li, Z.-K., Liu, X.-M., Xu, N.-Y. \& Du, J.-F. Experimental Realization of a Quantum Support Vector Machine. \textit{Phys. Rev. Lett.} \textbf{114}, 140504 (2015).

\bibitem{Gao2018} Gao, J. \textit{et al.} Experimental Machine Learning of Quantum States. \textit{Phys. Rev. Lett.} \textbf{120}, 240501 (2018).


\bibitem{Chatfield2004} Chatfield,C. The analysis of time series, sixth edition, New York, Chapman and Hall CRC (2004).

\bibitem{Fedrizzi2007} Fedrizzi,  A. \textit{et al.}  A wavelength-tunable fiber-coupled source of narrowband entangled photons. \textit{Opt. Express} \textbf{15}, 15377-15386 (2007).

\bibitem{James2001} James, D. F. V., Kwiat, P. G., Munro, W. J. \& White, A. G. Measurement of qubits. \textit{Phys. Rev. A} \textbf{64}, 052312 (2001).

\bibitem{Wootters1998} Wootters, W. K. Entanglement of formation of an arbitrary state of two qubits. \textit{Phys. Rev. Lett.} \textbf{80}, 2245 (1998).

\bibitem{Ranchin2016} Ranchin, A. Quantum Go, \textit{preprint arXiv:}1603.04751, (2016).

\bibitem{Robson1983} Robson, J. M. The complexcity of Go, in \textit{Information Processing 83: Proceedings of the IFIP Congress}, 413–417(1983)

\bibitem{Johanson2013} Johanson, M. Measuring the Size of Large No-Limit Poker Games, \textit{preprint arXiv:}1302.7008 (2013).

\bibitem{Zha2019} Zha, D. \textit{et al.} RLCard: A Toolkit for Reinforcement Learning in Card Games, \textit{preprint arXiv:}1910.04376, (2019).

\bibitem{Kim1996} Kim, T., Fiorentino, M. \& Wong, F. N. C. Phase-stable source of polarisation entangled photons using a polarisation Sagnac interferometer. \textit{Phys. Rev. A.} \textbf{73}, 012316 (2006).

\bibitem{Delpy1988} Delpy, D.T. Estimation of optical pathlength through tissue from direct time of flight measurement. \textit{Physics in Medicine and Biology} \textbf{33}, 1433-1422 (1988).

\end{thebibliography}

\begin{thebibliography}{}
\end{thebibliography}
\end{document}